\documentclass[aps,twocolumn,showpacs]{revtex4}
\usepackage{graphicx}
\usepackage{amsfonts}
\usepackage{amssymb}
\usepackage{amsbsy}
\usepackage{amsmath}
\usepackage{mathrsfs}
\usepackage{latexsym}
\usepackage{natbib}
\usepackage{bm}
\usepackage{subfigure} 
\usepackage{color}
\usepackage{wasysym}

\def\kr{\kappa}

\def\ksg{\mathrm{\varkappa}}

\def\rh{r_H}
\def\rstar{r_{\star}}

\def\scriplus{\mathscr{I}^{+}}
\def\scriminus{\mathscr{I}^{-}}

\def\observerminus{\mathbb{O}^{-}}
\def\observerplus{\mathbb{O}^{+}}

\DeclareMathOperator{\sech}{sech}

\begin{document}

\title{The Hawking effect and the bounds on greybody factor for higher 
dimensional Schwarzschild black holes}

\author{Subhajit Barman}
\email{sb12ip007@iiserkol.ac.in}

\affiliation{Department of Physical Sciences, 
Indian Institute of Science Education and Research Kolkata,
Mohanpur - 741 246, WB, India}
 

\date{\today}

\begin{abstract}

In this work, we have considered a $n-$dimensional Schwarzschild-Tangherlini 
black hole spacetime with massless minimally coupled free scalar fields in its 
bulk and $3-$brane. The bulk scalar field equation is separable using the higher 
dimensional spherical harmonics on $(n-2)-$sphere. First, using the Hamiltonian 
formulation with the help of the recently introduced near-null coordinates we 
have obtained the expected temperature of the Hawking effect, identical for both 
bulk and brane localized scalar fields. Second, it is known that the spectrum of 
the Hawking effect as seen at asymptotic future does not correspond to a perfect 
black body and it is properly represented by a greybody distribution. We have 
calculated the bounds on this greybody factor for the scalar field in both bulk 
and $3-$brane. Furthermore, we have shown that these bounds predict a decrease 
in the greybody factor as the spacetime dimensionality $n$ increases and suggest 
that for a large number of extra dimensions the Hawking quanta is mostly emitted 
in the brane.

\end{abstract}

\maketitle

\section{Introduction}\label{Introduction}

In 1975 Hawking \cite{hawking1975} predicted that semi classically black holes 
can also radiate particles, which is known as the Hawking effect. The Hawking 
effect is one of the most remarkable results of quantum field theory in curved 
spacetime, considered to be the cornerstone in understanding the black hole 
thermodynamics \cite{Bardeen:1973gs, Bekenstein:1973ur, Bekenstein:1974ax}. 
Usually Hawking effect is understood using Bogoliubov transformation between 
ingoing and outgoing field modes, which are described in terms of the null 
coordinates. Therefore, to provide a Hamiltonian based derivation of the Hawking 
effect one faces the basic hurdle that these null coordinates do not account for 
the dynamics of the spacetime and the construction of a non trivial field 
Hamiltonian is not possible using these coordinates. To overcome this difficulty 
in our previous work \cite{Barman:2017fzh} we have introduced a set of near-null 
coordinates which possess spacelike and timelike properties, and enables one to 
formulate the Hamiltonians for the study of the Hawking effect.

It is believed that inclusion of extra spatial dimensions in a spacetime have 
the potential of solving the hierarchy problem \cite{ArkaniHamed:1998rs, 
ArkaniHamed:1998nn, Antoniadis:1998ig, Kanti:2004nr, Randall:1999ee, 
Randall:1999vf}, which is related to the issue concerning the huge difference 
between the gravitational scale and Electro-Weak scale. Although experimental 
observation of the higher dimensional microscopic black holes are not yet 
verified in the latest state of the art particle colliders (LHC), these black 
holes remain interesting arenas to venture in for their enthralling properties 
coming from the extra dimensions. \vspace{0.0cm}

In this work we are going to consider a $n-$dimensional spherically symmetric 
static black hole spacetime, which is a higher dimensional generalization for 
the usual Schwarzschild black hole, better known as the 
Schwarzschild-Tangherlini \cite{Tangherlini1963} spacetime. Then we have 
considered scalar field both in the Bulk and in $3-$brane. The introduction of 
extra spatial dimensions changes the linear relation between the radius of the 
event horizon and the mass of the black hole, and it also has a signature in 
the surface gravity of the event horizon. Then the temperature corresponding to 
the Hawking effect gets modified. In this work we are going to provide a 
Hamiltonian based derivation of the Hawking effect in this $n-$dimensional 
Schwarzschild-Tangherlini black hole spacetime using the near-null coordinates 
\cite{Barman:2017fzh, Barman:2017vqx, Barman:2018ina}.

Furthermore, the spectrum of the Hawking radiation as seen by an asymptotic 
observer is not a complete black body distribution and it is properly described 
by a greybody distribution. The greybody factor is obtained from the 
transmission amplitude as the field modes pass from near horizon region to an 
asymptotic observer through the effective potential, created due to the 
spacetime geometry.
Estimation of this greybody factor is a very difficult job and often utilises 
various approximations like evaluating the greybody factors particularly in 
asymptotically low or high frequency regimes \cite{Harmark:2007jy, 
Keshet:2007be, Kim:2007gj, Rocha:2009xy, Gonzalez:2010vv, Gonzalez:2010ht, 
Kanti:2014dxa, CiprianA.Sporea:2019ary, Panotopoulos:2018pvu}. Sometimes, one is 
forced to take the extremal limit to evaluate these quantities 
\cite{Cvetic:1997uw, Cvetic:2009jn, Li:2009zzf}. Otherwise, one can take 
numerical approaches to estimate these greybody factors \cite{Harris:2003eg, 
Rocha:2009xy, Catalan:2014ama, Becar:2014aka, Dong:2015qpa, Pappas:2016ovo, 
Gray:2015pma, Abedi:2013xua}. On the other hand one can also evaluate the bounds 
on these greybody factors \cite{Visser:1998ke, Boonserm:2008qf, 
Boonserm:2009mi}, which have the advantages of providing analytical results even 
for intermediate frequency regimes and for all angular momentums 
\cite{Boonserm:2008zg, Ngampitipan:2012dq, Ngampitipan:2013sf, Boonserm:2013dua, 
Boonserm:2014rma, Boonserm:2014fja, Boonserm:2017qcq}.
We are going to estimate this bound on the greybody factor for scalar fields 
both in the bulk and in $3-$brane of the Schwarzschild-Tangherlini black hole 
spacetime. It is generally believed that for large number of extra dimensions, 
most of the energy corresponding to the Hawking effect is radiated in the brane 
\cite{Emparan:2000rs, Harris:2003eg, Chen:2008ra}. We intend to verify this 
phenomena from the bound on the greybody factors and also understand the 
dependence of this bound on the spacetime dimensionality $n$  
\cite{Creek:2006ia}.

In section \ref{ST-spacetime} we start with an introduction to the 
Schwarzschild-Tangherlini black hole spacetime. Then we consider a massless 
minimally coupled free scalar field both in the bulk and also in the $3-$brane 
of the higher dimensional spacetime. It is observed that the field equation in 
the bulk is separable in terms of the radial and angular coordinates using the 
higher dimensional spherical harmonics on $(n-2)-$dimensional sphere. We also 
notice that both the bulk and brane scalar field action can be transformed to a 
$(1+1)$ dimensional reduced form. In the subsequent section 
\ref{canonical-formulation} we first define the near-null coordinates and then 
using these coordinates provide the formulation for the canonical derivation of 
the Hawking effect. In section \ref{HawkingST} we give a Hamiltonian based 
derivation of the Hawking effect in the Schwarzschild-Tangherlini black hole 
spacetime. In the subsequent section \ref{Bound_GBFactor} we evaluate the bounds 
on the greybody factor, and discuss the results in section \ref{discussion}.

\section{Schwarzschild-Tangherlini black hole spacetime}\label{ST-spacetime}

Higher dimensional black holes have their own relevance in brane world models 
and string theory where large extra spatial dimensions can exist, 
\cite{Kaloper:2000jb, Starkman:2000dy, Starkman:2001xu, Cvetic:1997uw, 
Cvetic:1996xz}. The statistical counting of entropy was first provided in string 
theory for a 5-dimensional black hole \cite{Strominger:1996sh}. In this section 
we are going to discuss the Schwarzschild-Tangherlini (ST) black hole spacetime 
\cite{Tangherlini1963}, which is a higher dimensional generalization of the 
$(3+1)-$dimensional Schwarzschild black holes.

\subsection{Metric and horizon of the ST black hole}

The generalization of the four dimensional Schwarzschild black holes to higher 
dimensions, say $n-$dimensions, is described by the Schwarzschild-Tangherlini 
line-element \cite{Singh:2017vfr, Pereniguez:2018xng, Emparan:2008eg, 
Cardoso:2005vb, Harris:2003eg, Cardoso:2005mh, Ngampitipan:2012dq, 
Kanti:2004nr}, given by
\begin{equation}\label{eq:ST_metric_bulk}
 ds^2 = -f(r)dt^2 + \frac{1}{f(r)}dr^2 + r^2 d\Omega^2_{n-2}~,
\end{equation}
which satisfies the $n-$dimensional vacuum Einstein's equations. Here 
$d\Omega^2_{n-2}$ denotes the metric on a $(n-2)-$dimensional unit sphere. The 
function $f(r)$ is given by
\begin{equation}\label{eq:ST_fr}
 f(r) = 1-\left(\frac{\rh}{r}\right)^{n-3}~,
\end{equation}
where
\begin{equation}\label{eq:ST_rH}
 \rh = \left[\frac{16GM}{(n-2)\tilde{\eta}_{n-2}}\right]^{\frac{1}{n-3}}~~; 
~~\tilde{\eta}_{n-2} = 
\frac{2\pi^{\frac{n-1}{2}}}{\Gamma\left(\frac{n-1}{2}\right)}~,
\end{equation}
and $G=1/M_{Pl}^{n-2}$ is the $(n-2)$ dimensional Newton constant. Here 
$\tilde{\eta}_{n-2}$ denotes the volume of the unit $(n-2)$ dimensional sphere. 
We note that $r=\rh$ describes the position of the event horizon. From Eqn. 
(\ref{eq:ST_metric_bulk}) and (\ref{eq:ST_fr}) we also observe that when $n=4$ 
the metric reduces to the $4-$dimensional Schwarzschild metric. On the other 
hand from Eqn.(\ref{eq:ST_rH}) we notice that the radius of the event horizon is 
not linearly related to the mass of the black hole for $n>4$. 

The metric on the  $3-$brane is obtained by fixing the angular coordinates, 
which represent the excess dimensions in addition to the $(3+1)-$dimensional 
spacetime. In our case we have fixed the coordinates $\theta_{i}=\pi/2$ with $i$ 
from $1$ to $(n-4)$. The line element on the brane is expressed as
\begin{equation}\label{eq:ST_metric_brane}
 ds^2 = -f(r)dt^2 + \frac{1}{f(r)}dr^2 + r^2 d\Omega^2_{2}~,
\end{equation}
where $d\Omega^2_{2}$ denotes the line element on a unit two sphere, with 
$f(r)$ and $\rh$ given by the same previous expressions. In ST black hole 
spacetime the tortoise coordinate $\rstar$ is obtained from
\begin{eqnarray}
 \frac{d\rstar}{dr} &=& \frac{1}{f(r)}\nonumber\\ 
 ~&=& \frac{n-2}{2(n-3)} + \frac{\rh}{(n-3)(r-\rh)} + g'(r-\rh)~,
\end{eqnarray}
where $g'(r-\rh)$ is a polynomial function of $(r-\rh)$ such that $g'(0)=0$ and 
it is assumed to represent the derivative of $g(r-\rh)$ with respect to $r$. 
Then with suitable choice of integration constant the tortoise coordinate 
becomes
\begin{equation}\label{eq:rstar1}
 \rstar = \frac{n-2}{2(n-3)}r + 
\frac{\rh}{(n-3)}\ln{\left(\frac{r}{\rh}-1\right)} + g(r-\rh)~.
\end{equation}
The expression of the tortoise coordinate from Eqn. (\ref{eq:rstar1}) is crucial 
to get a relation between the spatial near-null coordinates for the two 
observers corresponding to the Hawking effect. Furthermore, this relation in the 
limit $r\to\rh$ determines the spectrum of the Hawking effect.

\subsection{Reduced scalar field action}

We consider a massless minimally coupled free scalar field $\Phi(x)$ in 
$n-$dimensional ST black hole spacetime with the action given by
\begin{equation}\label{eq:scalar-action-full}
S_{\Phi} = \int d^{n}x \left[ -\frac{1}{2} \sqrt{-g} g^{\mu \nu} 
\nabla_{\mu}\Phi(x) \nabla_{\nu}\Phi(x) \right]~.
\end{equation}
We first consider the scalar field in the bulk and the corresponding line 
element is taken from Eqn. (\ref{eq:ST_metric_bulk}). In $n-$dimensional 
spacetime the coordinates $x^{\mu}$ are $\left(t, r, 
\theta_{1},...,\theta_{n-2}\right)$ where $0\le\theta_{n-2}\le2\pi$ and 
$0\le\theta_{i}\le\pi$, for $i=1,...,n-3$. In ST black hole spacetime the 
determinant of the metric tensor $g$ provides
\begin{equation}
 \sqrt{-g} = r^{n-2} 
\sin^{n-3}{\theta_{1}}\sin^{n-4}{\theta_{2}}...\sin{\theta_{n-3}}~.
\end{equation}
The components of the inverse metric are $g^{00}=-1/f(r)$, $g^{11}=f(r)$, 
$g^{22}=1/r^2$, $g^{33}=1/\left(r^2\sin^2{\theta_{1}}\right)$,..., 
$g^{n-1,n-1}=1/\left(r^2\sin^2{\theta_{1}}...\sin^2{\theta_{n-3}}\right)$. We 
assume the field decomposition $\Phi(x) = \sum_{l,m}\tilde{\varphi}_{lm}(t,r) ~ 
Y^{n}_{lm}(\Omega)$, where $Y^{n}_{lm}(\Omega)$ is the spherical harmonics on 
the $(n-2)-$dimensional sphere. Then the action from Eqn. 
(\ref{eq:scalar-action-full}) becomes
\begin{eqnarray}\label{eq:ST_action1}
 &&S_{\Phi} = \frac{1}{2}\int 
dtdrd\theta_{1}...d\theta_{n-2}~\sqrt{-g}\left[ \frac{1}{f}(\partial_{t}\Phi)^2 
- f(\partial_{r}\Phi)^2\right. \nonumber\\
&&-\left.\frac{1}{r^2}\left\{ (\partial_{\theta_{1}}\Phi)^2 + 
\frac{(\partial_{\theta_{2}}\Phi)^2}{\sin^2{\theta_{1}}}+...+\frac{
(\partial_ { \theta_{n-2}}\Phi)^2}{\sin^2{ 
\theta_{1}}...\sin^2{ 
\theta_{n-3}}} \right\} \right]\nonumber\\
&&~~~~=~ \sum_{l,m}\int dtd\rstar~\frac{r^{n-2}}{2}\left[ 
(\partial_{t}\tilde{\varphi}_{lm})^2 - 
(\partial_{\rstar}\tilde{\varphi}_{lm})^2 \right.\nonumber\\
~&&~~~~~~~~~~~~~~~~~~~~~~~~~~~~~~\left.+\frac{\mathcal{A}^{n}_{lm}}{r^2}~\tilde{
\varphi } _ { lm } ^2 \right]~,
\end{eqnarray}
where the angular integrals are carried out and $\mathcal{A}^{n}_{lm}$ is the 
eigen-value of the $n-$dimensional spherical harmonics equation 
\cite{Frye:2012jj}. We further assume 
$\tilde{\varphi}_{lm}=\varphi_{lm}(t,r)/r^{\frac{n-2}{2}}$ then the action 
reduces to $S_{\Phi} = \sum_{lm} S_{lm}$, where
\begin{eqnarray}\label{eq:ST_action2_bulk}
 S_{lm} &=& \frac{1}{2}\int dtd\rstar\left[ (\partial_{t}\varphi_{lm})^2 - 
(\partial_{\rstar}\varphi_{lm})^2   \right. \nonumber\\
~&+& \left.  \frac{(n-2)f}{r}\varphi_{lm} 
\partial_{\rstar}\varphi_{lm}
-\frac{f\varphi_{lm}^2}{r^2}\left\{f\left(\tfrac{n-2}{2}
\right)^2 - \mathcal{A}^{n}_{lm}\right\} \right]~.\nonumber\\
\end{eqnarray}
For a brane localized scalar field the line element is taken from Eqn. 
(\ref{eq:ST_metric_brane}) and action for the scalar field is defined by the 
same Eqn. (\ref{eq:scalar-action-full}) with $n=4$. Then performing a similar 
field decomposition, however this time with respect to the spherical harmonics 
on two sphere and $\tilde{\varphi}_{lm}=\varphi_{lm}(t,r)/r$, we shall obtain 
the action after the angular integrals are carried out, as 
\begin{eqnarray}\label{eq:ST_action2_brane}
 S_{lm} &=& \frac{1}{2}\int dtd\rstar\left[ (\partial_{t}\varphi_{lm})^2 - 
(\partial_{\rstar}\varphi_{lm})^2   \right. \nonumber\\
~&+& \left.  \frac{2f}{r}\varphi_{lm} 
\partial_{\rstar}\varphi_{lm}
-\frac{f\varphi_{lm}^2}{r^2}\left\{f - \mathcal{A}^{4}_{lm}\right\} 
\right]~.\nonumber\\
\end{eqnarray}
Now in the asymptotic regions, i.e. near scriminus $\scriminus$ and scriplus 
$\scriplus$ we have $r\to\infty$, on the other hand at the event horizon 
$f(\rh)=0$. Then in asymptotic regions and in regions near the horizon the 
above actions from Eqn. (\ref{eq:ST_action2_bulk}) and Eqn. 
(\ref{eq:ST_action2_brane}) simplifies to
\begin{equation}\label{eq:ST_action3}
 S_{lm} \simeq \frac{1}{2}\int dtd\rstar\left[ (\partial_{t}\varphi_{lm})^2 - 
(\partial_{\rstar}\varphi_{lm})^2 \right]~,
\end{equation}
which represents the action corresponding to a $(1+1)$ dimensional flat 
Minkowski spacetime. This reduction enables one to utilize the techniques of 
field theory from standard flat spacetime to these specific regions of the ST 
black hole spacetime. The ST metric is independent of time, thus the spacetime 
is time translation invariant and one may express the field mode 
$\varphi_{lm}(t,\rstar) = e^{i\omega t}\psi_{lm}(\rstar)$. Then from Eqn. 
(\ref{eq:ST_action2_bulk}) and Eqn. (\ref{eq:ST_action2_brane}) one can find 
out the corresponding equation of motions as
\begin{equation}\label{eq:STregge-wheeler}
 \partial^{2}_{\rstar}\psi_{lm} + \left[ \omega^2 - \mathbb{V}(r) 
\right]\psi_{lm} = 0~,
\end{equation}
with the potential for the field in the bulk, given by
\begin{equation}\label{eq:ST_potential_bulk}
 \mathbb{V}(r) = f(r)\left[ \frac{(n-2)(n-4)f(r)}{4r^2}+ \frac{(n-2)f'(r)}{2r} - 
\frac{\mathcal{A}^{n}_{lm}}{r^2} \right]~,
\end{equation}
and the potential for the brane localized scalar field
\begin{equation}\label{eq:ST_potential_brane}
 \mathbb{V}(r) = f(r)\left[\frac{f'(r)}{r} - \frac{\mathcal{A}^{4}_{lm}}{r^2} 
\right]~,
\end{equation}
where $f'(r)$ denotes differentiation of $f(r)$ with respect to $r$. We also 
mention that the eigen-value of the $n-$dimensional spherical harmonics 
\cite{Cardoso:2002pa, Frye:2012jj} is given by $\mathcal{A}^{n}_{lm}=-l(l+n-3)$.

\subsection{Number density of Hawking quanta at $\scriplus$}

The field equation corresponding to the action (\ref{eq:ST_action3}) has 
solutions given by
\begin{equation}\label{eq:field_sol_asymp}
 \varphi_{lm}\sim\frac{1}{\sqrt{2\pi\omega}}e^{-i\omega(t\pm\rstar)}~.
\end{equation}
Therefore the field modes near past null infinity, near future null infinity 
and near the horizon have characteristics of plane waves. In this work and as 
also discussed in the original work by Hawking the black hole is assumed to be 
formed through the collapse of matters, where the initial spacetime in the past 
is described by a Minkowski metric. On the other hand the final metric is 
denoted by respective black hole spacetime. It is generally accepted that this 
dynamical nature of the spacetime metric generates the notion of particle 
creation. However, for a derivation of the Hawking radiation it is not necessary 
to understand the details of the collapse. In particular, according to Hawking 
the Bogoliubov transformation between the modes at past null infinity 
$\scriminus$ and the outgoing modes which have just escaped the creation of the 
black hole provide the Planckian distribution of the Hawking effect. For a 
massless minimally coupled free scalar field described by the action 
(\ref{eq:scalar-action-full}) the corresponding number density of Hawking quanta 
is
\begin{equation}\label{eq:ND_general}
 N_{\omega} = \frac{1}{e^{2\pi\omega/\ksg_{H}}-1}~,
\end{equation}
where $\ksg_{H}$ is the surface gravity at the event horizon and $\omega$ 
corresponds to the frequency of the wave mode. The field modes then pass through 
the black hole spacetime to be observed by an asymptotic observer near future 
null infinity $\scriplus$. We have seen from Eqn. (\ref{eq:STregge-wheeler}) 
that there is a potential barrier between the horizon and the spatial infinity 
at $\scriplus$. Then because of this potential barrier, spectrum of the Hawking 
radiation observed at $\scriplus$ would also get modified and the above 
mentioned black body distribution transforms to a Greybody distribution. The 
transmission probability through the potential barrier denotes the greybody 
factor and the number density corresponding to the Hawking quanta at $\scriplus$ 
becomes
\begin{equation}\label{eq:ND_general_GB}
 N_{\omega} = \frac{\Gamma\left(\omega\right)}{e^{2\pi\omega/\ksg_{H}}-1}~,
\end{equation}
where $\Gamma\left(\omega\right)$ denotes the greybody factor, which 
generally depends on the frequency $\omega$ of the outgoing modes and on the 
black hole parameters.

\section{Canonical formulation}\label{canonical-formulation}

Usually, the Bogoliubov transformation between ingoing and outgoing field modes, 
expressed in terms of the null coordinates $v$ and $u$, generates the Hawking 
effect. However, these null coordinates do not describe the dynamics of the 
system, and one cannot construct true matter field Hamiltonian out of them. To 
overcome this difficulty we consider the recently introduced near-null 
coordinates \cite{Barman:2017fzh}. Using these near-null coordinates the 
Hamiltonians for the two observers $\observerminus$ and $\observerplus$, which 
are located respectively near the past null infinity $\scriminus$ and near the 
horizon, are formed. We mention that two different time-independent metrics 
correspond to the spacetimes for these two observers, with the one near the 
event horizon developed through the collapse of a matter shell. This dynamical 
nature of the spacetime, which evolves from being $n-$dimensional Minkowski to 
$n-$dimensional Schwarzschild, directly governs particle creation in curved 
spacetime. We note that in \cite{Barman:2017fzh}, the observer $\observerplus$ 
was assumed to be stationed near $\scriplus$, which was reasonable as there we 
did not consider the greybody factors and the spectrum of the Hawking effect 
near the horizon was regarded to be the spectrum seen by an asymptotic observer.

\subsection{The observers $\observerminus$ and $\observerplus$}

\subsubsection{Near-null coordinates}

Here we define a set of \emph{near-null} coordinates by slightly deforming the 
null coordinates \cite{Barman:2017fzh, Barman:2017vqx, Barman:2018ina}. In 
particular for observer $\observerminus$ we define the near-null coordinates as
\begin{equation}\label{NearNullCoordinatesMinus}
\tau_{-} = t - (1-\epsilon)\rstar ~~;~~ \xi_{-} = -t - (1+\epsilon)\rstar  ~,
\end{equation}
where $\epsilon\gg\epsilon^2$ is considered to be a small parameter. Similarly, 
we define the near-null coordinates for observer $\observerplus$, as
\begin{equation}\label{NearNullCoordinatesPlus}
\tau_{+} = t + (1-\epsilon)\rstar ~~;~~ \xi_{+} = -t + (1+\epsilon)\rstar  ~.
\end{equation}
We note that timelike characteristics of coordinates $\tau_{-}$ and $\tau_{+}$ 
are preserved in the whole range $0<\epsilon<2$ of the parameter $\epsilon$, 
though for convenience this parameter is considered to be small in our case.

\subsubsection{Field Hamiltonian}

We are considering the black hole to be formed by the collapse of matters with 
the spacetime for the observer $\observerminus$ near $\scriminus$ being flat 
Minkowskian. Then for the past observer $\observerminus$, the $(1+1)$ 
dimensional reduced spacetime is described by the Minkowski metric $ ds^2 = 
-dt^2 + d\rstar^2 = -dt^2 + dr^2$. On the other hand, for the future observer 
$\observerplus$ near the horizon, the black hole is already formed, and the 
$(1+1)$ dimensional reduced spacetime can be expressed by the metric $ ds^2 =  
-dt^2 + d\rstar^2$. Using the near-null coordinates 
(\ref{NearNullCoordinatesPlus}) and (\ref{NearNullCoordinatesMinus}), the 
invariant line-elements for observers $\observerplus$ and $\observerminus$ 
become
\begin{equation}\label{NearNullMetricPM}
ds^2_{\pm} = \tfrac{\epsilon}{2} [ - d\tau_{\pm}^2 + 
d\xi_{\pm}^2 +\tfrac{2}{\epsilon} d\tau_{\pm} d\xi_{\pm} ] 
\equiv \tfrac{\epsilon}{2}~ g^{0}_{\mu\nu}dx_{\pm}^{\mu} dx_{\pm}^{\nu}  ~ ,
\end{equation}
where flat metric $g^{0}_{\mu\nu}$ is conformally transformed. Then for both of 
the observers the reduced scalar field action (\ref{eq:ST_action3}) can be 
expressed as 
\begin{equation}\label{ReducedScalarAction2DFlatMinus}
S_{\varphi} =  \int d\tau_{\pm}  d\xi_{\pm} \left[-\tfrac{1}{2} \sqrt{-g^{0}} 
g^{0\mu\nu} \partial_{\mu}\varphi \partial_{\nu} \varphi \right]  ~.
\end{equation}
From Eqn. (\ref{NearNullMetricPM}) we observe that the \emph{lapse function} $N 
= 1/\epsilon$, \emph{shift vector} $N^1 = 1/\epsilon$ and determinant of the 
spatial metric $q = 1$. Then the scalar field Hamiltonians for observers 
$\observerplus$ and $\observerminus$ can be written as
\begin{equation}\label{ScalarHamiltonianOpm}
H_{\varphi}^{\pm} = \int d\xi_{\pm}\; \tfrac{1}{\epsilon}\left[\left\{ 
\tfrac{1}{2} \Pi^2  + \tfrac{1}{2}  (\partial_{\xi_{\pm}}\varphi)^2 \right\} 
+ \Pi~ \partial_{\xi_{\pm}} \varphi \right] ~.
\end{equation}
From this Eqn. (\ref{ScalarHamiltonianOpm}) we observe that when $\epsilon=0$, 
the Hamiltonians become ill-defined, which signifies the necessity of having 
near-null coordinates to construct the Hamiltonians. Using Hamilton's equation, 
the field momentum $\Pi$ can be expressed as
\begin{equation}\label{FieldMomentumPM}
\Pi(\tau_{\pm},\xi_{\pm}) = \epsilon\;\partial_{\tau_{\pm}}\varphi - 
\partial_{\xi_{\pm}}\varphi ~,
\end{equation}
which satisfies a Poisson bracket with the field $\varphi$, as
\begin{equation}\label{PoissonBracketPM}
 \{\varphi(\tau_{\pm},\xi_{\pm}), \Pi(\tau_{\pm},\xi_{\pm}')\} = 
\delta(\xi_{\pm} - \xi_{\pm}') ~.
\end{equation}

\subsubsection{Fourier modes}

We consider finite fiducial boxes in the intermediate steps of calculations for 
both of the observers to avoid dealing with formally divergent spatial volumes 
$\int d\xi_{\pm}\sqrt{q}$, where $\sqrt{q}=1$. The finite volumes are given by
\begin{equation}\label{SpatialVoumePM}
V_{\pm} = \int_{\xi_{\pm}^L}^{\xi_{\pm}^R} d\xi_{\pm}\sqrt{q} = {\xi_{\pm}^R} - 
{\xi_{\pm}^L}  ~.
\end{equation}
Subsequently, we define the respective Fourier modes of the scalar field and the 
conjugate momentum for the observers $\observerplus$ and $\observerminus$ as
\begin{eqnarray}
\varphi(\tau_{\pm},\xi_{\pm}) &=& \tfrac{1}{\sqrt{V_{\pm}}}\sum_{k} 
\tilde{\phi}^{\pm}_{k} 
e^{i k \xi_{\pm}}  ~,
\nonumber \\
\Pi(\tau_{\pm},\xi_{\pm}) &=&  \tfrac{1}{\sqrt{V_{\pm}}} \sum_{k} \sqrt{q}~ 
\tilde{\pi}^{\pm}_{k} 
e^{i k \xi_{\pm}} ~,
\label{FourierModesDefinitionPM}
\end{eqnarray}
where $\tilde{\phi}^{\pm}_{k} = \tilde{\phi}^{\pm}_{k} (\tau_{\pm})$, 
$\tilde{\pi}^{\pm}_{k} = \tilde{\pi}^{\pm}_{k} (\tau_{\pm})$ are the 
complex-valued mode functions. The finite volumes of the fiducial boxes lead to 
the definition of Kronecker delta and Dirac delta as $\int d\xi_{\pm}\sqrt{q}  
e^{i(k-k')\xi_{\pm}} = V_{\pm} \delta_{k,k'}$ and $\sum_k 
e^{ik(\xi_{\pm}-\xi_{\pm}')} = V_{\pm} \delta(\xi_{\pm}-\xi_{\pm}')/\sqrt{q}$ .
The definition of these two deltas together imply $k\in \{k_s\}$ where 
$k_s=2\pi s/V_{\pm}$ with $s$ being a nonzero integer. These definitions help us 
to express the scalar field Hamiltonians (\ref{ScalarHamiltonianOpm}) in terms 
of the Fourier modes as $H_{\varphi}^{\pm} = \sum_{k} \tfrac{1}{\epsilon}
\left(\mathcal{H}_{k}^{\pm} + \mathcal{D}_{k}^{\pm}\right)$ where the 
Hamiltonian densities and diffeomorphism generators are
\begin{equation}\label{FourierHamiltonianPM}
\mathcal{H}_{k}^{\pm} = \tfrac{1}{2} \tilde{\pi}^{\pm}_{k}  
\tilde{\pi}^{\pm}_{-k} + \tfrac{1}{2} k^2 \tilde{\phi}^{\pm}_{k} 
\tilde{\phi}^{\pm}_{-k} ~,
\end{equation}
and
\begin{equation}\label{DiffeomorphismGenerator}
\mathcal{D}_{k}^{\pm} =  
 -\tfrac{i k}{2} ( \tilde{\pi}^{\pm}_{k} \tilde{\phi}^{\pm}_{-k} -
 \tilde{\pi}^{\pm}_{-k} \tilde{\phi}^{\pm}_{k} )  ~,
\end{equation}
respectively. The corresponding Poisson brackets are
\begin{equation}\label{FourierPoissonBracketPlus}
\{\tilde{\phi}^{\pm}_{k}, \tilde{\pi}^{\pm}_{-k'}\} = \delta_{k,k'} ~.
\end{equation}
The number density corresponding to Hawking quanta will be obtained from the 
expectation value of the Hamiltonian density operator for the field modes of the 
observer $\observerplus$ in the vacuum state of observer $\observerminus$.

\subsection{Relation between Fourier modes}
    
It can be shown that the Fourier components of the scalar field and its 
conjugate momentum for the two observers maintain relations among 
themselves, analogous to the Bogoliubov transformation, given by
\begin{equation}\label{FieldModesRelation}
\tilde{\phi}^{+}_{\kr} = \sum_{k} \tilde{\phi}^{-}_{k} F_{0}(k,-\kr) ~;~
\tilde{\pi}^{+}_{\kr} =  \sum_{k} \tilde{\pi}^{-}_{k} F_{1}(k,-\kr)  ~,
\end{equation}
where the Fourier modes are considered on fixed spatial hyper-surfaces. The 
coefficient functions $F_{n}(k,\kr)$, where $n=0,1$, are obtained from the 
relation $\varphi(\tau_{-},\xi_{-}) = \varphi(\tau_{+},\xi_{+})$, as the field 
is scalar, and the relation $\Pi(\tau_{+},\xi_{+}) = (\partial \xi_{-}/\partial 
\xi_{+}) \Pi(\tau_{-},\xi_{-})$ \cite{Barman:2017fzh} between the field 
momentums. We note that for $k,\kr >0$ the coefficient functions 
$F_{n}(-k,-\kr)$ are analogous to the Bogoliubov mixing coefficients 
$\beta_{\omega\omega'}$ whereas $F_{n}(k,-\kr)$ are analogous to the Bogoliubov 
coefficients $\alpha_{\omega\omega'}$ of \cite{hawking1975}. The expression of 
these coefficient functions $F_{n}(k,\kr)$ are given by
\begin{equation}\label{FFunctionGeneral}
F_{n}(k,\kr) = \frac{1}{\sqrt{V_{-} V_{+}}} \int d\xi_{+} 
\left(\tfrac{\partial \xi_{-}}{\partial \xi_{+}}\right)^n
~e^{i k \xi_{-} + i \kr \xi_{+}} ~.
\end{equation}
Using this mathematical expression of the coefficient functions one can obtain a 
relation
\begin{equation}\label{F0F1Relation}
F_{1}(\pm k,\kr) = \mp \left(\tfrac{\kr}{k}\right) F_{0}(\pm k,\kr) ~,
\end{equation}
which indicates evaluating only $F_{0}(\pm k,\kr)$ is sufficient for the 
subsequent analysis.

\subsection{Consistency condition and relation between Hamiltonian densities and 
diffeomorphism generators}

The coefficient functions $F_0(\pm k,\kr)$ satisfy a relation among themselves, 
which results from the simultaneous satisfaction of the two different Poisson 
brackets $\{\tilde{\phi}^{-}_{k},\tilde{\pi}^{-}_{-k'}\} = \delta_{k,k'}$ and 
$\{\tilde{\phi}^{+}_{\kr},\tilde{\pi}^{+}_{-\kr'}\} = \delta_{\kr,\kr'}$. Using 
the Eqn. (\ref{F0F1Relation}) this relation can be expressed as
\begin{equation}\label{PoissonBracketConsistencyCondition}
\mathbb{S}_{-}(\kr) - \mathbb{S}_{+}(\kr) =\sum_{k>0} \frac{\kr}{k} 
\left[|F_{0}(-k,\kr)|^2-|F_{0}(k,\kr)|^2\right] = 1 ~.
\end{equation}
This relation is analogous to the consistency condition between Bogoliubov 
coefficients \cite{Bhattacharya:2013tq}. Using relations 
(\ref{FieldModesRelation}) and (\ref{F0F1Relation}) one can express the 
Hamiltonian density $\mathcal{H}_{\kr}^{+}$ for the observer $\observerplus$ in 
terms of the Hamiltonian density $\mathcal{H}_{k}^{-}$ of the observer 
$\observerminus$ as 
\begin{equation}\label{ModesHamiltonianRelations0}
\mathcal{H}_{\kr}^{+} = h_{\kr}^1 + \sum_{k>0} \left(\frac{\kr}{k}\right)^2 
\left[\right|F_{0}(-k,\kr)|^2 + |F_{0}(k,\kr)|^2] ~\mathcal{H}_k^{-}  ~,
\end{equation}
where 
$h_{\kr}^1 =  \sum_{k\neq k'} (\kr^2/2 k k') F_{0}(k,-\kr) F_{0}(-k',\kr) 
\{\tilde{\pi}^{-}_{k} \tilde{\pi}^{-}_{-k'} + kk' \tilde{\phi}^{-}_{k} 
\tilde{\phi}^{-}_{-k'} \}$. $h_{\kr}^1$ being linear in $\phi^{-}_{k}$ and 
its conjugate momentum, the vacuum expectation value of its quantum counterpart 
vanishes. Similarly, the diffeomorphism generators of the two observers can be 
related as
\begin{equation}\label{ModesDiffeomorphismRelations0}
\mathcal{D}_{\kr}^{+} = d_{\kr}^1 + \sum_{k>0} \left(\frac{\kr}{k}\right)^2 
\left[\right|F_{0}(-k,\kr)|^2 + |F_{0}(k,\kr)|^2] ~\mathcal{D}_k^{-}  ~,
\end{equation}
where  $ d_{\kr}^1 = \sum_{k\neq k'} (i\kr^2/2k)~ \{F_{0}(-k,\kr) 
F_{0}(k',-\kr)~ \tilde{\pi}^{-}_{-k}  \tilde{\phi}^{-}_{k'} - F_{0}(k,-\kr) 
F_{0}(-k',\kr)~ \tilde{\pi}^{-}_{k} \tilde{\phi}^{-}_{-k'} \}$ which is also 
linear in field mode and its conjugate momentum.

\subsection{Fock quantization and the vacuum state}

We define real-valued field modes out of the complex valued Fourier modes 
$\tilde{\phi}_{k}=\tilde{\phi}^{*}_{-k}$ by suitably choosing the real-valued 
parts \cite{Hossain:2010eb,Barman:2017fzh}, such that this newly defined filed 
modes are all independent. Then the Hamiltonian density represent simple 
harmonic oscillators as
\begin{equation}\label{RealHamiltonian}
\mathcal{H}_{k}^{\pm} = \frac{1}{2} \pi^{2}_{k}
+ \frac{1}{2} k^2 \phi^{2}_{k} 
~,~~~\{\phi^{2}_{k},\pi^{2}_{k'}\}=\delta_{k,k'}~,
\end{equation}
and the diffeomorphism generator vanish $\mathcal{D}_k^{-} = 0$, where 
$\phi_{k}$ and $\pi_{k}$ are the redefined real-valued field modes.
Now we restrict ourselves with the modes where $k,\kr > 0$ and for the 
considered massless scalar field the mode frequency can be identified as $\omega 
= k$. For $k^{th}$ oscillator mode the energy spectrum is given by 
$\hat{\mathcal{H}}_{k}^{-}|n_k\rangle = (\hat{N}_{k}^{-}+\tfrac{1}{2}) k 
|n_k\rangle = (n+\tfrac{1}{2}) k |n_k\rangle $ where $\hat{N}_{k}^{-}$ is the 
number operator, $|n_k\rangle$ represent eigen-states with integer eigenvalues. 
We realize the Hawking effect by computing the expectation value of the 
Hamiltonian density operator $\hat{\mathcal{H}}_{\kr}^{+} \equiv 
(\hat{N}_{\kr}^{+}+\tfrac{1}{2})\kr$ corresponding to the observer 
$\observerplus$ in the vacuum state $|0_{-}\rangle = \Pi_{k} |0_k\rangle$ 
corresponding to the observer $\observerminus$. Therefore, the expectation value 
of the number density operator corresponding to the Hawking quanta of frequency 
$\omega = \kr$, using the Eqn. (\ref{PoissonBracketConsistencyCondition}) and 
Eqn. (\ref{ModesHamiltonianRelations0}), can be expressed as
\begin{equation}\label{NumberDensityEVGeneral} 
N_{\omega} = N_{\kr} \equiv \langle 0_{-}|\hat{N}_{\kr}^{+}|0_{-}\rangle 
=  \mathbb{S}_{+}(\kr)   ~,
\end{equation}
where we have used $\langle 0_k|\hat{\phi}_{k} |0_k\rangle = 0$ and $\langle 
0_k|\hat{\pi}_{k} |0_k\rangle = 0$.

\section{The Hawking effect for Schwarzschild-Tangherlini black 
holes}\label{HawkingST}

In actual derivation of the Hawking effect it was shown that there is a 
logarithmic relation between the ingoing and outgoing null coordinates in a 
black hole spacetime, which is crucial to get the thermal distribution of the 
Hawking effect. In canonical formulation we are going to obtain a similar 
relation between the spatial near-null coordinates of the two observers, which 
produces the number density for Hawking quanta in an analogous manner.

\subsection{Relation between spatial coordinates $\xi_{-}$ and $\xi_{+}$}

In order to establish the relation between the coordinates $\xi_{-}$ 
and $\xi_{+}$, following \cite{Barman:2017fzh}, we consider a pivotal 
point $\xi_{-}^0$ on a $\tau_{-} = constant$ hyper-surface. A spacelike 
interval on this hyper-surface can be written as
\begin{equation}\label{ScriMinusXiMinuxInterval}
(\xi_{-} - \xi_{-}^0)_{|\tau_{-}} = 2(\rstar^0-\rstar)_{|\tau_{-}} =
2(r^0-r)_{|\tau_{-}} \equiv \Delta ~, 
\end{equation}
where $r^0$ is a pivotal value corresponding to $\xi_{-}^0$. In deriving Eqn. 
(\ref{ScriMinusXiMinuxInterval}) we have used fact that for the observer 
$\observerminus$ the spacetime was Minkowskian. In a similar manner using the 
expression of the tortoise coordinate from Eqn. (\ref{eq:rstar1}) we can express 
a spacelike interval on a $\tau_{+} = constant$ hyper-surface as
\begin{eqnarray}\label{ScriMinusXiPlusInterval}
(\xi_{+} - \xi_{+}^0)_{|\tau_{+}} 
&=& \frac{n-2}{2(n-3)}\Delta + \frac{1}{\ksg_{H}} \ln \left(1 + 
\frac{\Delta}{\Delta_{0}}\right)\nonumber\\ 
~&&+2~g\left(\frac{\Delta+\Delta_{0}}{2}\right)
- 2~g\left(\frac{\Delta_{0}}{2}\right)~,
\end{eqnarray}
where $\Delta_{0} \equiv 2(r^0 - r_{H})_{|\tau_{+}}$ and $\ksg_{H}$ is the 
surface gravity at the event horizon. Further, we have identified the interval 
$2(r - r^0)_{|\tau_{+}}$ as $\Delta$ using \emph{geometric optics 
approximation}. We choose the pivotal values $\xi_{-}^0 = \Delta_{0}$ and 
$\xi_{+}^0 = \frac{n-2}{2(n-3)}\xi_{-}^0 + \frac{1}{\ksg_{H}} \ln (\ksg_{H} 
\xi_{-}^0)+ 2~g(\xi_{-}^0/2)$. These choices lead to the 
relation
\begin{equation}\label{eq:near-null-ralation}
 \xi_{+} = \frac{n-2}{2(n-3)}\xi_{-} + \frac{1}{\ksg_{H}} \ln (\ksg_{H} 
\xi_{-})+ 2~g\left(\frac{\xi_{-}}{2}\right) ~.
\end{equation}
The modes that give rise to the Hawking radiation, travel out from the region 
very close to the horizon and for them $\ksg_{H}\xi_{-} \ll 1$. Consequently 
for these modes, the relation (\ref{eq:near-null-ralation}) can be approximated 
as
\begin{equation}\label{eq:final-near-null-relation}
  \xi_{+} \approx \frac{1}{\ksg_{H}} \ln (\ksg_{H}\xi_{-})  ~.
\end{equation}
We note from the Eqn. (\ref{eq:final-near-null-relation}) that the full domain 
of the coordinate $\xi_{+}$ is $(-\infty,\infty)$ whereas it is $(0,\infty)$ 
for $\xi_{-}$ \emph{i.e.} the domains are the same as implied by the Eqn. 
(\ref{eq:near-null-ralation}). However, as mentioned earlier, we shall restrict
ourselves within a finite fiducial box during the intermediate steps in our 
analysis.

\subsection{Evaluation of coefficient functions $F_{0}(\pm k,\kr)$}

From the Eqns. (\ref{PoissonBracketConsistencyCondition}) and 
(\ref{NumberDensityEVGeneral}) we observe that the consistency condition and the 
number density of the Hawking quanta both require the expression  of 
$F_{0}(\pm k,\kr)$, which can be written as
\begin{equation}\label{F0NonExtremal}
F_{0}(\pm k,\kr) = \int \frac{d\xi_{-}}{\sqrt{V_{-} V_{+}}}
\frac{e^{\pm i k \xi_{-} + i (\kr/\ksg_{H})\ln 
(\ksg_{H}\xi_{-})}}{\ksg_{H}\xi_{-}}  ~.
\end{equation}
The integrand being oscillatory in nature, the coefficient function 
$F_{0}(k,\kr)$ (\ref{F0NonExtremal}) is formally divergent. In order to 
regulate this integral we introduce the standard `$i\delta$' regulator, with 
small $\delta >0$, as follows
\begin{eqnarray}\label{F0NonExtremalRegulated}
 F_{0}^{\delta}(\pm k,\kr) &=& 
\int \frac{d\xi_{-}}{\sqrt{V_{-} V_{+}}}
(\ksg_{H}\xi_{-})^{-1} 
~ e^{\pm i k \xi_{-}}  \nonumber \\ 
&& ~ \times ~ e^{(\delta + i\kr/\ksg_{H} )\ln (\ksg_{H}\xi_{-})} ~.
\end{eqnarray}
In the limit $\delta\to 0$, the regulated expression $F_{0}^{\delta}(\pm k,\kr)$ 
 reduces to $F_{0}(\pm k,\kr)$. By introducing variables  $b_0 = (\delta + 
i\kr/\ksg_{H})$ and $\xi = k ~\xi_{-}$, we can express regulated coefficient 
function as
\begin{eqnarray}\label{F0NonExtremalEvaluated}
 F_{0}^{\delta}(\pm k,\kr) &=& 
 \frac{(\ksg_{H}/k)^{b_0}}{\ksg_{H} \sqrt{V_{-} V_{+}}} 
\int d\xi ~e^{\pm i\xi} ~\xi^{b_0-1}\nonumber\\  
~&=& \frac{(\ksg_{H}/k)^{b_0} ~\Gamma(b_0)}{\ksg_{H} \sqrt{V_{-} V_{+}}} 
e^{\pm i{b_0} \pi /2} ~,
\end{eqnarray}
where $\Gamma(b_0)$ is the Gamma function. Given the fiducial box has a finite 
volume, we have added two boundary terms $\Delta I^{L} = \int_0^{\xi^L} d\xi 
e^{\pm i\xi} \xi^{b_0-1}$ and $\Delta I^{R} = \int_{\xi^R}^{\infty} d\xi 
e^{\pm i\xi} 
\xi^{b_0-1}$ to make the Gamma function complete. Both of these terms vanish 
when one removes the volume regulators by taking the limit $\xi^L \equiv 
(k~\xi_{-}^L)\to 0$ and $\xi^R \equiv 
(k~\xi_{-}^R)\to\infty$. We note an useful property
\begin{equation}\label{F0F0Relation}
F_{0}^{\delta}(-k,\kr) = e^{\pi(\kr/\ksg_{H} - 
i\delta)}~F_{0}^{\delta}(k,\kr)  ~.
\end{equation}
Eqn. (\ref{F0F0Relation}) shows that these coefficient functions satisfy a 
relation analogous to the relation between Bogoliubov coefficients from 
\cite{hawking1975}.

\subsection{Consistency condition}          

The Eqn. (\ref{F0NonExtremalEvaluated}) together with the relation 
$k := k_s = (2\pi s/V_{-})$ leads to
\begin{equation}\label{SplusNonExtremal1}
\mathbb{S}^{\delta}_{+}(\kr) = \frac{\kr ~|\Gamma(b_0)|^2 
e^{-\pi\kr/\ksg_{H}}}
{\ksg_{H}^{2-2\delta} (2\pi)^{1+2\delta}} ~ 
\left(\frac{\zeta(1+2\delta)}{V_{-}^{-2\delta} V_{+}} \right) ~,
\end{equation}
where $\zeta(1+2\delta) = \sum_{s=1}^{\infty} s^{-(1+2\delta)}$ is the 
\emph{Riemann zeta function}. Furthermore, the Eqn. (\ref{F0F0Relation}) 
implies that $\mathbb{S}^{\delta}_{-}(\kr) = e^{2\pi\kr/\ksg_{H}} 
~\mathbb{S}^{\delta}_{+}(\kr)$. Given $\zeta(1)$ is divergent, it is clear that  
in order to keep the term $\mathbb{S}^{\delta}_{\pm}$ finite one needs to 
remove volume regulators $\xi_{-}^L$ and $\xi_{-}^R$ along with the integral 
regulator $\delta$. To find the required dependency among the regulators, we use 
the regulated expression (\ref{F0NonExtremalEvaluated}) such that the 
consistency condition (\ref{PoissonBracketConsistencyCondition}) becomes
\begin{equation}\label{PoissonBracketConsistencyCondition2}
\frac{\sinh(\pi\kr/\ksg_{H})}
{\pi~(\kr/\ksg_{H})^{-1} |\Gamma(b_0)|^{-2}} = 
\frac{(\ksg_{H}V_{+}) (2\pi/\ksg_{H}V_{-})^{2\delta} }{\zeta(1+2\delta)} ~.
\end{equation}
Using Gamma function identity $\Gamma(z)\Gamma(1-z) = \pi/\sin\pi z$, zeta 
function identity $\lim_{\delta\to0}[\delta~ \zeta(1+\delta)] = 1$ and the 
Eqn. (\ref{eq:final-near-null-relation}) one can show that the consistency 
condition demands $\ksg_{H} \xi_{-}^L \sim e^{-1/2\delta}$, i.e. the volume 
regulator $\xi_{-}^L$ and integral regulator $\delta$ should be varied together. 
Once this limit is taken other volume regulator $\xi_{-}^R$ drops off from the 
expression of $\mathbb{S}^{\delta}_{+}(\kr)$.

\subsection{Number density of Hawking quanta}

Therefore, the expectation value of the number density operator 
(\ref{NumberDensityEVGeneral}) for a Schwarzschild-Tangherlini black hole as 
seen by observer $\observerplus$ is
\begin{equation}\label{NumberDensityEVNonExtremal} 
N_{\kr} = \langle\hat{N}_{\kr=\omega}\rangle 
= \frac{1}{e^{2\pi\omega/\ksg_{H}} - 1} ~. 
\end{equation}
which represents a blackbody distribution at the Hawking temperature $T_H \equiv 
\ksg_{H}/(2\pi k_B) = (n-3)/(4\pi k_B r_H)$ \cite{Feng:2015jlj, Feng:2015exa, 
Matyjasek:2014dfa}. Clearly, the Hawking temperature for ST black hole depends 
both on its mass $M$, represented by $r_H$ and the dimensionality of the 
spacetime $n$. We also want to mention that this number density as observed by 
an asymptotic observer will be represented by a greybody distribution, given by
\begin{equation}\label{NumberDensityNonExtremalPhysical}
N_{\omega} = \frac{\Gamma(\omega)}{e^{2\pi\omega/\ksg_{H}} - 1} ~.
\end{equation}
In the next section we center our attention on this greybody factor.

\section{Bounds on the greybody factor}\label{Bound_GBFactor}

As discussed in our previous section the spectrum of Hawking quanta as seen by 
an asymptotic future observer is described by a greybody distribution, which is 
modified from the blackbody distribution as the wave modes pass through the 
black hole spacetime. Any physical observer that intends to detect the Hawking 
radiation is expected to be stationed at a far away distance from the black hole 
event horizon. Then the estimation of the greybody factor becomes important to 
obtain a general form of the Hawking spectrum to these observers. In literature 
researchers have introduced diverse technicalities \cite{Cardoso:2005vb, 
Cardoso:2005mh, Harmark:2007jy, Grain:2005my, Fernando:2004ay, Ida:2002ez, 
Cvetic:1998sp, Cvetic:1997ap, Klebanov:1997cx, Gray:2015xig} to estimate these 
greybody factors. In this work we are going to provide the bounds on these 
greybody factors for ST black holes. Unlike other approximate methods, these 
bounds can be estimated for all frequencies, field angular momentums and 
spacetime dimensions.

\subsection{Bounds}

\begin{figure}
 \includegraphics[width=0.8\linewidth]{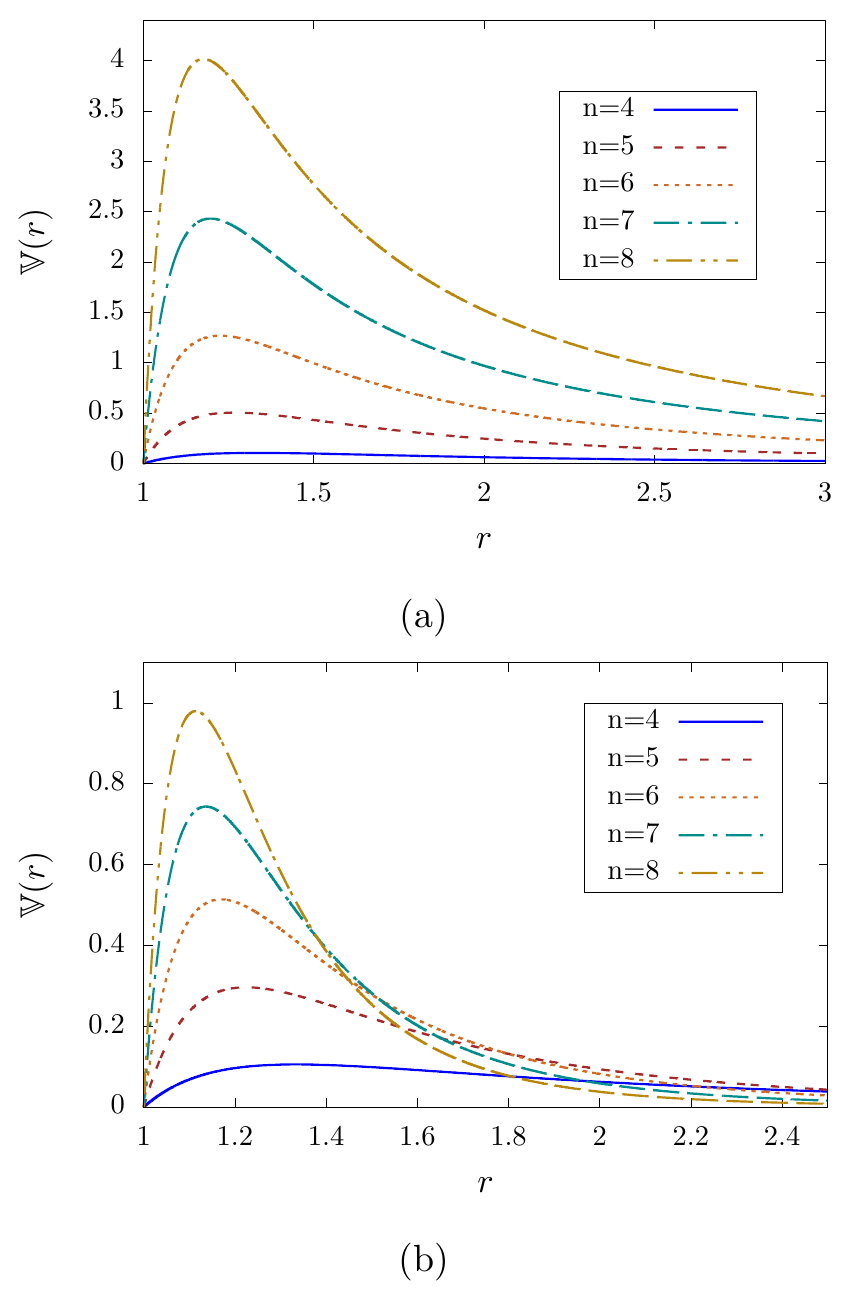}
 \caption{(a) Plot of the potential $\mathbb{V}(r)$ corresponding to a scalar 
field in the bulk with respect to the radial coordinate $r$ for $l=0$, $r_{H}=1$ 
and different dimensionality $n$. (b) Plot of the potential $\mathbb{V}(r)$ 
corresponding to a brane-localized scalar field with respect to the radial 
coordinate $r$ for $l=0$, $r_{H}=1$ and different dimensionality $n$.}
 \label{fig:potential_ST}
\end{figure}

To obtain the bound on the greybody factor in ST black hole spacetime we 
consider the equation of motion from Eqn. (\ref{eq:STregge-wheeler}) with the 
potentials from Eqn. (\ref{eq:ST_potential_bulk}) and Eqn. 
(\ref{eq:ST_potential_brane}). Eigen-values of the $n-$dimensional spherical 
harmonics are given by $\mathcal{A}^{n}_{lm}=-l(l+n-3)$. In 
FIG.\ref{fig:potential_ST} we have depicted the potential as a varying function 
of the radial distance $r$ for different spacetime dimensionality $n$ 
corresponding to scalar field in bulk and in brane. We further use these 
potentials to get the general bounds on the greybody factor following 
\cite{Visser:1998ke, Boonserm:2008qf}. These bounds on the greybody factor, also 
introduced in \cite{Boonserm:2008zg}, can be expressed as
\begin{equation}\label{eq:bound-GBF}
 \Gamma(\omega) \geq 
\sech^2{\left\{\int_{-\infty}^{\infty}\varrho~~ d\rstar\right\}}~,
\end{equation}
where 
\begin{equation}\label{eq:rho-GBF}
 \varrho = \frac{\sqrt{(h')^2+(\omega^2-\mathbb{V}-h^2)^2}}{2h}~.
\end{equation}
Here $\mathbb{V}$ represents the potential, $h\equiv h(\rstar)$ and 
$h(\rstar)>0$, which is an arbitrary function satisfying $h(-\infty) = h(\infty) 
= \omega$. We consider two particular functional forms of $h$ 
\cite{Boonserm:2008zg} and obtain the resulting bound on the greybody factor in 
ST black hole spacetime.

\vspace{0.3cm}

\paragraph{case I:} Here we consider the situation when $h=\omega$. For this 
case $\varrho = \mathbb{V}/2\omega$ and the bound is given by
\begin{equation}\label{eq:GBF-case1-1}
 \Gamma(\omega) \geq 
\sech^2{\left\{\frac{1}{2\omega}\int_{\rh}^{\infty} 
\frac{\mathbb{V}(r)}{f(r)}dr \right\}}~.
\end{equation}
%

\begin{figure}
 \includegraphics[width=0.8\linewidth]{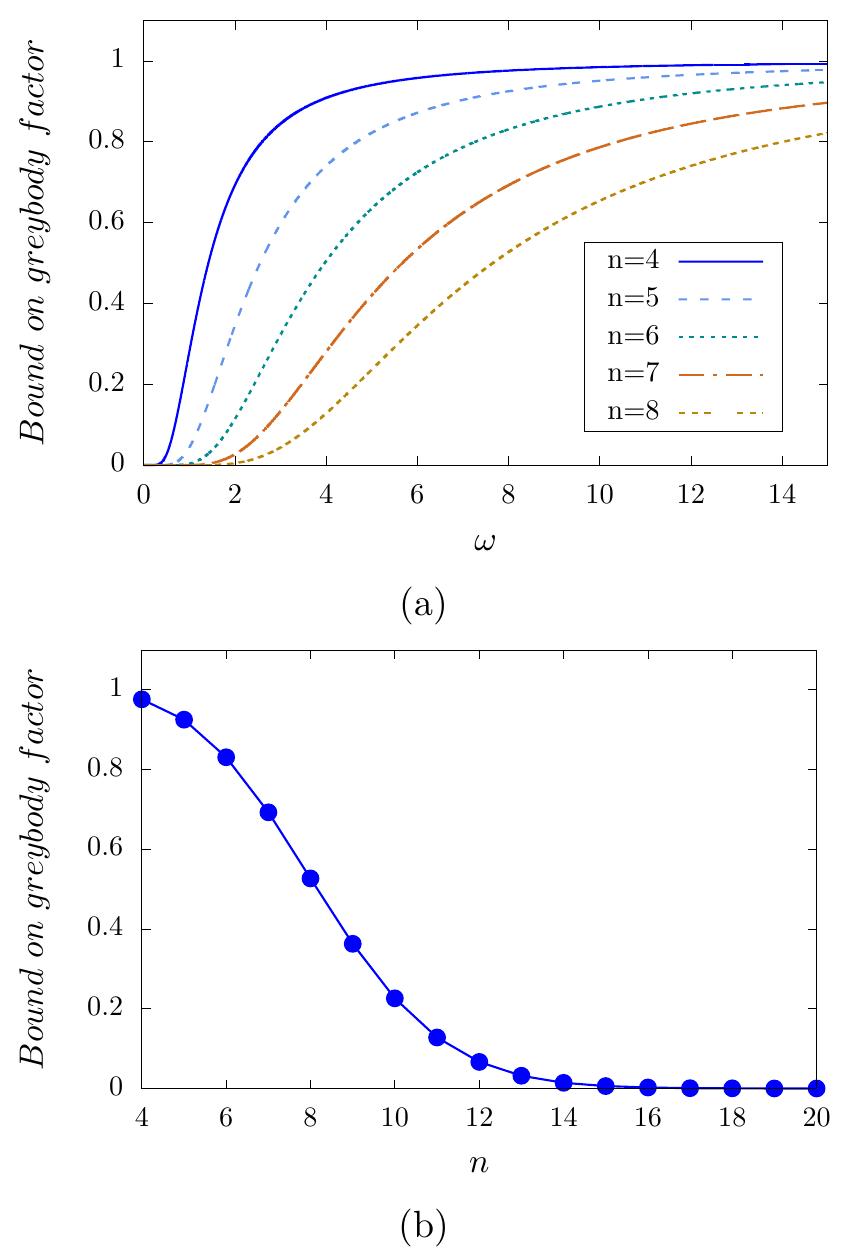}
 \caption{(a) Here we plot the bound on the greybody factor with respect to the 
frequency $\omega$ for $l=1$, $\rh=1$ and different values of spacetime 
dimensionality $n$. (b) Here we provide the plot of the bound on the greybody 
factor with respect to varying spacetime dimension $n$ for $l=1$, $\rh=1$. The 
frequency $\omega=8$ is chosen conveniently. In both of these cases the 
considered potential corresponds to a scalar field in the bulk and the 
particular assumption on $h$ is $h=\omega$.}
 \label{fig:c1_w&n_dep_Bulk}
\end{figure}

\begin{figure}
 \includegraphics[width=0.8\linewidth]{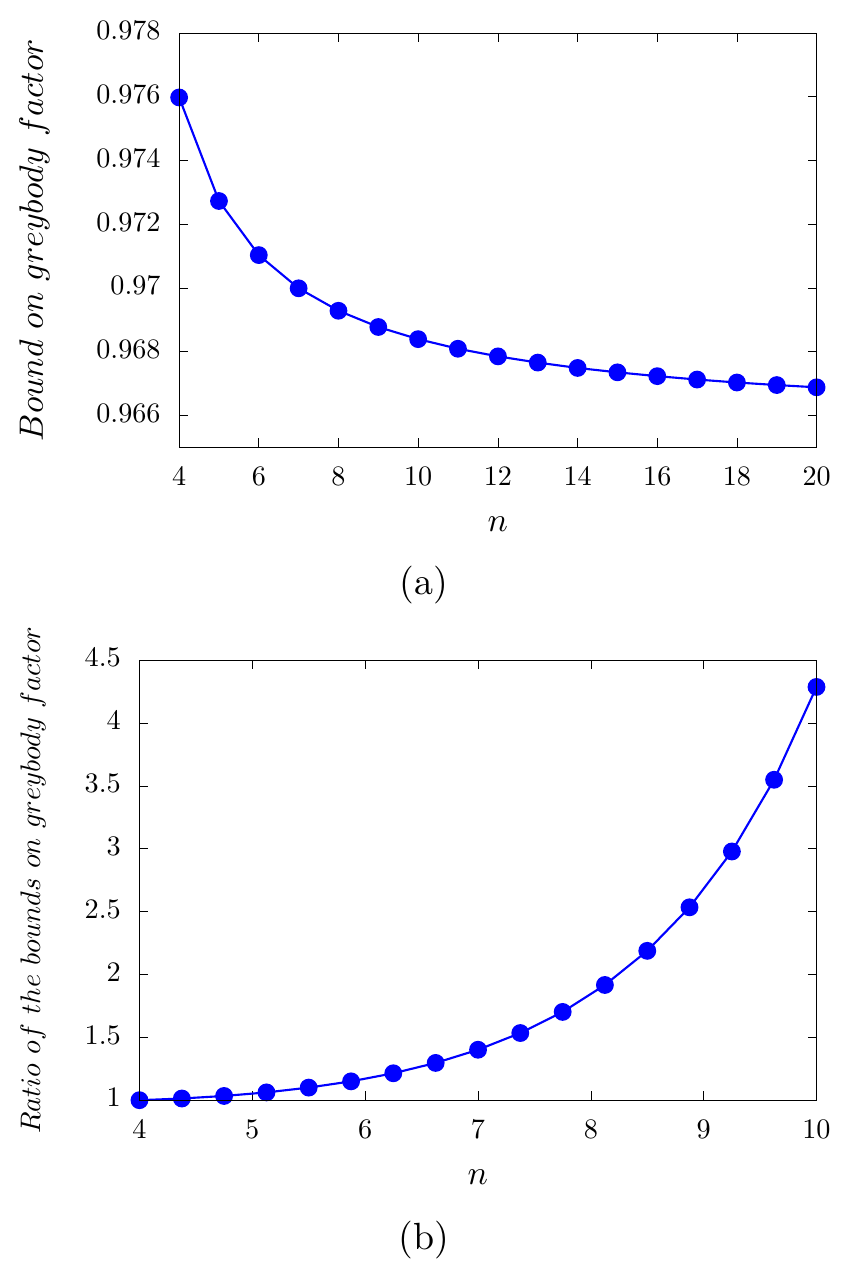}
 \caption{(a) Here we plot the bound on the greybody factor, corresponding to a 
brane-localized scalar field, with respect to different spacetime dimension $n$ 
for $l=1$, $\rh=1$. (b) The plot of the ratio of the bounds on the greybody 
factor, corresponding to the brane to bulk scalar field, with respect to 
varying dimension $n$ for $l=1$, $\rh=1$. For both of the cases the frequency 
$\omega=8$ is chosen conveniently and the particular assumption on $h$ is 
$h=\omega$.}
 \label{fig:c1_ndep_Br}
\end{figure}

First, we consider the scalar field in the bulk. Using the functional form of 
potential from Eqn. (\ref{eq:ST_potential_bulk}) one can get this bound 
evaluated to be
\begin{equation}\label{eq:GBF-case1-1_bulk}
  \Gamma(\omega) \geq 
\sech^2{\left\{\frac{6+4l^2+4l(n-3)+n^2-5n}{8\omega \rh} \right\}}~,
\end{equation}
which is non-zero, finite for any infinitesimally small $\omega$, i.e. for very 
small frequency also there is some transmission amplitude. When $\omega$ is 
large one can series expand the above quantity and the bound becomes
\begin{eqnarray}\label{eq:GBF_case1_hW_bulk}
    \Gamma(\omega) \geq 1-
\frac{\left[6+4l^2+4l(n-3)+n^2-5n\right]^2}{64\rh^2\omega^2}+ 
\mathcal{O}(\frac{1}{\omega^4})~.\nonumber\\
\end{eqnarray}

This result signifies that for very high frequency outgoing waves the potential 
barrier becomes almost transparent. We observe that for $n=4$ the above 
expression from Eqn. (\ref{eq:GBF-case1-1_bulk}) reproduces the bound obtained 
for $(3+1)$ dimensional Schwarzschild black hole \cite{Boonserm:2008zg}. In FIG. 
\ref{fig:c1_w&n_dep_Bulk} we have depicted the variation of this bound on the 
greybody factor with respect to the frequency $\omega$ for different values of 
the spacetime dimensionality $n$. We also observe that the bound decreases as 
$n$ increases, see FIG. \ref{fig:c1_w&n_dep_Bulk}. This feature, also mentioned 
in \cite{Boonserm:2008zg}, is a consequence of the fact that as $n$ increases 
the height of the potential barrier increases and thus the transmission 
probability should decrease.

On the other hand one can also consider the brane-localized scalar field and 
take the potential from Eqn. (\ref{eq:ST_potential_brane}) to obtain the bound 
as 
\begin{equation}\label{eq:GBF-case1-1_brane}
  \Gamma(\omega) \geq 
\sech^2{\left\{\frac{n+l(l+1)(n-2)-3}{2\omega \rh (n-2)} \right\}}~.
\end{equation}
One can observe that when $n=4$ this expression gives the bound obtained for 
$(3+1)$ dimensional Schwarzschild black hole. For large $\omega$ this bound can 
be series expanded to give 
\begin{eqnarray}\label{eq:GBF_case1_hW_brane}
    \Gamma(\omega) \geq 1-
\left[\frac{n+l(l+1)(n-2)-3}{2\rh\omega(n-2)}\right]^2+ 
\mathcal{O}\left(\frac{1}{\omega^4}\right)~,\nonumber\\
\end{eqnarray}
and similarly for large $n$ one gets,
\begin{eqnarray}\label{eq:GBF_case1_nH_brane}
    \Gamma(\omega) \geq 
\sech^2\left[\frac{1+l+l^2}{2\rh\omega}\right]+ 
\mathcal{O}\left(\frac{1}{n}\right)~.
\end{eqnarray}
From Eqn. (\ref{eq:GBF_case1_hW_brane}) one can notice that with the rise of 
frequency the transmission probability grows also for a brane-localized scalar 
field. On the other hand from Eqn.(\ref{eq:GBF-case1-1_bulk}) and Eqn. 
(\ref{eq:GBF_case1_nH_brane}) we observe that unlike the bulk case the lower 
bound on the greybody factor never goes to zero even for infinitely large $n$ 
for a scalar field in $3-$brane. From FIG. \ref{fig:c1_ndep_Br} we observe that 
this bound slowly decreases as $n$ increases for a brane localized scalar field 
compared to the field in bulk. Besides, in this particular figure we observe 
that the ratio of these bounds for brane to bulk is always greater than one and 
increases as the spacetime dimensionality increases. This suggests that for a ST 
black hole with large extra spatial dimensions most of the Hawking emission 
arrives to an asymptotic observer in $3-$brane.

\vspace{0.3cm}
\paragraph{case II:}In this part we consider the ansatz 
$h=\sqrt{\omega^2-\mathbb{V}}$ \cite{Boonserm:2008zg}. With this form of $h$ the 
expression of $\varrho$ from Eqn. (\ref{eq:rho-GBF}) gets simplified and the 
bound on the greybody factor can be represented by the integral
\begin{equation}\label{eq:GBF-case2-1}
  \Gamma(\omega) \geq 
\sech^2{\left\{\frac{1}{2}\int_{-\infty}^{\infty} 
\frac{|h'|}{h}d\rstar \right\}}~.
\end{equation}
After carrying out the integration we get 
\begin{equation}\label{eq:GBF-case2-2}
  \Gamma(\omega) \geq 
\sech^2{\left\{-\ln{\left(\frac{\sqrt{\omega^2-\mathbb{V}_{peak}}}{\omega}
\right)} 
\right\}}~,
\end{equation}
where $\mathbb{V}_{peak}$ represents the maximum value or the peak value of the 
potential $\mathbb{V}$. One can observe from this bound that it has meaning only 
when $\omega^2>\mathbb{V}_{peak}$. The region $\omega^2<\mathbb{V}_{peak}$ is 
classically forbidden and do not contribute to this particular calculation. The 
expression of the bound from Eqn. (\ref{eq:GBF-case2-2}) can be further 
simplified to
\begin{equation}\label{eq:GBF-case2-3}
 \Gamma(\omega) \geq 
1 - \frac{\mathbb{V}_{peak}^2}{(2\omega^2-\mathbb{V}_{peak})^2}~.
\end{equation}
%

\begin{figure}
 \includegraphics[width=0.8\linewidth]{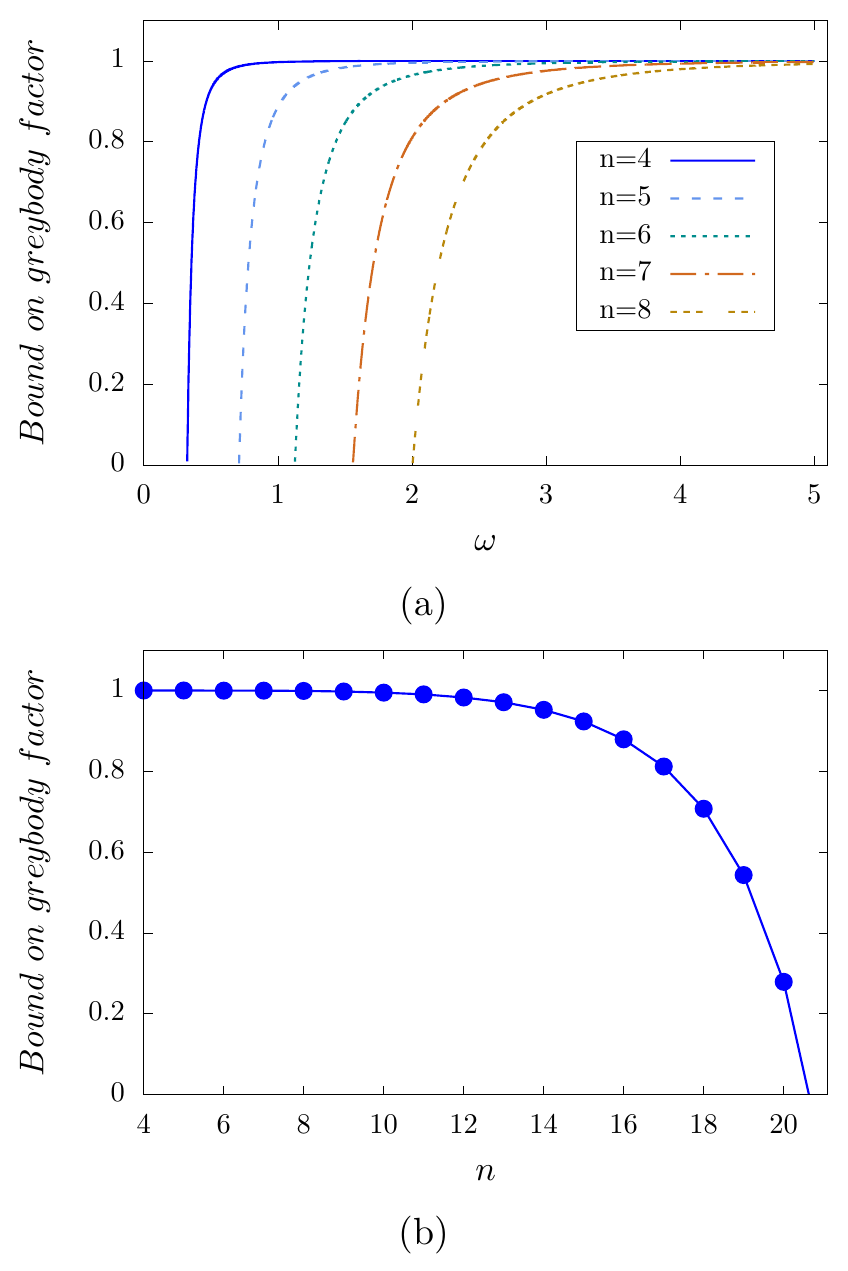}
 \caption{(a) Here we plot the bound on the greybody factor with respect to the 
frequency $\omega$ for $l=0$, $\rh=1$ and different values of spacetime 
dimension $n$. (b) The plot of the bound on the greybody factor with respect to 
varying spacetime dimensionality $n$ for $l=0$, $\rh=1$ and a conveniently 
chosen frequency $\omega=8$. Both of the plots correspond to a scalar field in 
the Bulk and the considered case is $h=\sqrt{\omega^2-\mathbb{V}}$.}
 \label{fig:c2_w&n_dep_Bulk}
\end{figure}

\begin{figure}
 \includegraphics[width=0.8\linewidth]{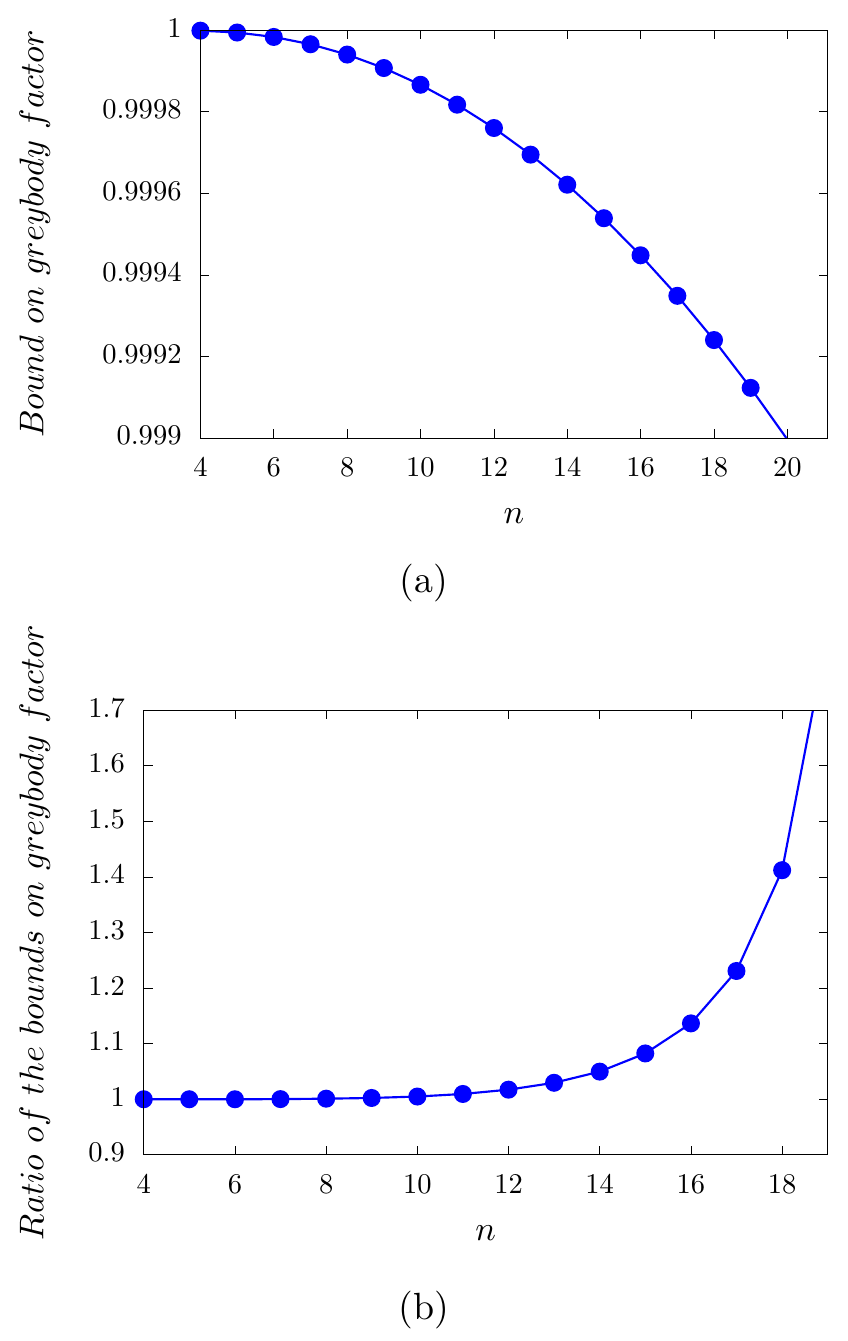}
 \caption{(a) Here we plot the bound on the greybody factor, corresponding to a 
brane-localized scalar field, with respect to different spacetime dimension $n$ 
for $l=1$, $\rh=1$. (b) The plot of the ratio of the bounds on the greybody 
factor, corresponding to the brane to bulk scalar field, with respect to 
varying dimension $n$ for $l=1$, $\rh=1$. For both of the cases the frequency 
$\omega=8$ is chosen conveniently and the particular assumption on $h$ is 
$h=\sqrt{\omega^2-\mathbb{V}}$.}
 \label{fig:c2_ndep_Br}
\end{figure}

To estimate this bound explicitly for certain frequency $\omega$, angular 
momentum $l$ and spacetime dimensionality $n$, one needs to evaluate the value 
of $r_{peak}$ which corresponds to the $\mathbb{V}_{peak}$ in each case. We have 
numerically done this particular computation for $l=0$ and also for varying $l$ 
for a scalar field in both bulk and in the brane. \vspace{0.05cm}

We have observed that when $l=0$ the plot given in FIG. 
\ref{fig:c2_w&n_dep_Bulk} shows similar characteristics as was obtained from the 
previous case for a scalar field in the bulk. We have seen that as $\omega$ 
increases the bound increases and potential barrier becomes more transparent 
FIG. \ref{fig:c2_w&n_dep_Bulk}, i.e. high frequency wave modes travel much 
smoothly than the low frequency modes through the potential barrier. Here also 
as the spacetime dimensionality $n$ increases the bound decreases, see FIG. 
\ref{fig:c2_w&n_dep_Bulk}. However in this particular case one can not expect to 
get the bound for arbitrarily small frequency $w$, as 
$\omega^2<\mathbb{V}_{peak}$ was excluded from the calculation. One can also 
analytically obtain the value of $r_{peak}$ corresponding to a scalar field in 
the bulk, when $l=0$, and it is given by
\begin{equation}\label{eq:rpeak_c2_l0_Blk}
 r_{peak} = \rh~ 
\left[\frac{n-1+(n-3)\sqrt{4n-7}}{2~(n-2)^2}\right]^{-\frac{1}{n-3}},
\end{equation}
where for few values of the spacetime dimension $n$ we have tabulated the 
corresponding $r_{peak}$ and $\mathbb{V}_{peak}$ as
\begin{center}
 \begin{tabular}{ | c | c | c | }
 \hline
  $n=4$ & $r_{peak}=4\rh/3$ & $\mathbb{V}_{peak}=0.105/\rh^2$\\
 \hline
  $n=5$ & $r_{peak}=1.267\rh$ & $\mathbb{V}_{peak}=0.505/\rh^2$\\
 \hline
  $n=6$ & $r_{peak}=1.226\rh$ & $\mathbb{V}_{peak}=1.269/\rh^2$\\
 \hline
  $n=7$ & $r_{peak}=1.197\rh$ & $\mathbb{V}_{peak}=2.432/\rh^2$\\
 \hline
  $n=8$ & $r_{peak}=1.176\rh$ & $\mathbb{V}_{peak}=4.017/\rh^2$\\
 \hline
 \end{tabular}
\end{center}
where we explicitly observe that as $n$ increases $r_{peak}$ gets nearer to the 
horizon and the value of $\mathbb{V}_{peak}$ increases for a given horizon 
radius $\rh$. For a brane localized scalar field we have, when $l=0$, the 
expression of $r_{peak}$ as
\begin{equation}\label{eq:rpeak_c2_l0_Br}
 r_{peak} = \rh\left[\frac{n-1}{2~(n-2)}\right]^{-\frac{1}{n-3}}.
\end{equation}
Some of the values of this $r_{peak}$ and corresponding $\mathbb{V}_{peak}$ 
for different spacetime dimensionality $n$ are tabulated below
\begin{center}
 \begin{tabular}{ | c | c | c | }
 \hline
  $n=4$ & $r_{peak}=4\rh/3$ & $\mathbb{V}_{peak}=0.105/\rh^2$\\
 \hline
  $n=5$ & $r_{peak}=1.225\rh$ & $\mathbb{V}_{peak}=0.296/\rh^2$\\
 \hline
  $n=6$ & $r_{peak}=1.170\rh$ & $\mathbb{V}_{peak}=0.514/\rh^2$\\
 \hline
  $n=7$ & $r_{peak}=1.136\rh$ & $\mathbb{V}_{peak}=0.744/\rh^2$\\
 \hline
  $n=8$ & $r_{peak}=1.114\rh$ & $\mathbb{V}_{peak}=0.980/\rh^2$\\
 \hline
 \end{tabular}
\end{center}
%

\begin{figure}
 \includegraphics[width=0.8\linewidth]{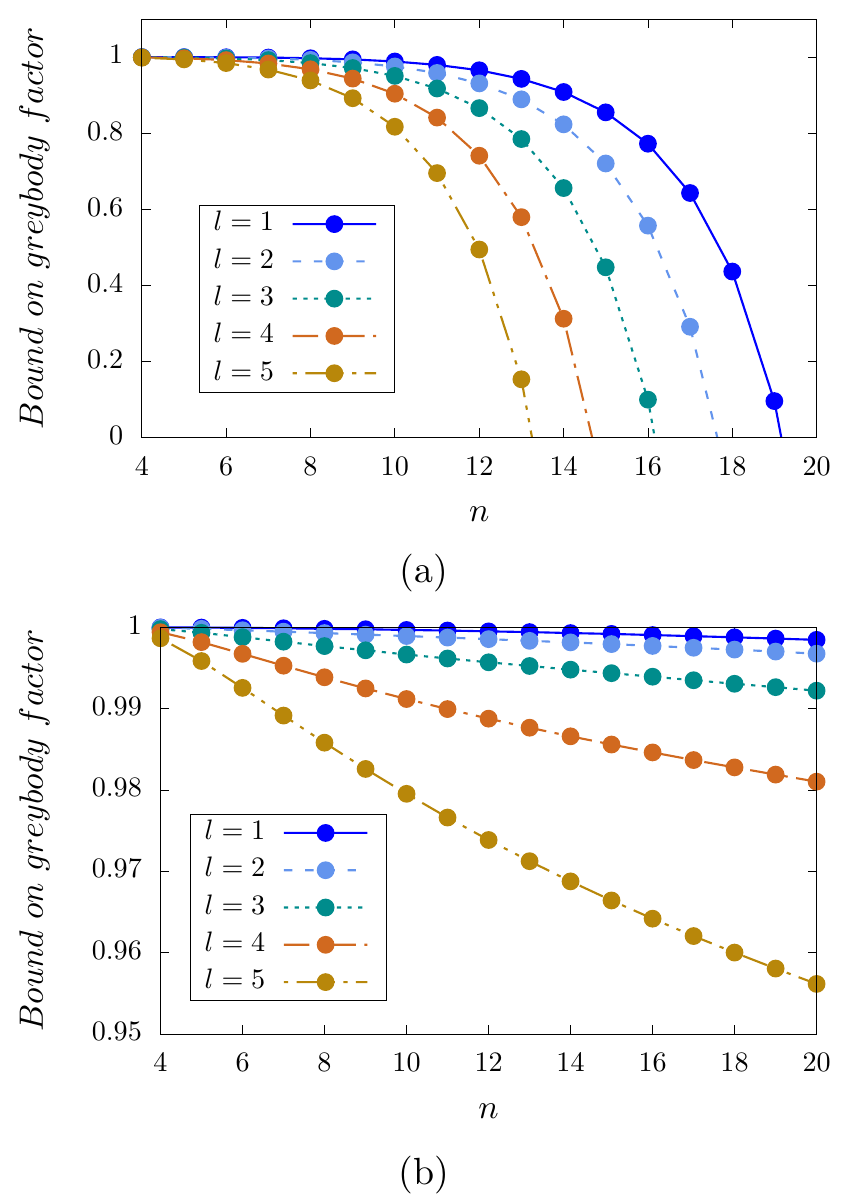}
 \caption{(a) Here we plot the bound on the greybody factor for a scalar field 
in Bulk with respect to varying spacetime dimensionality $n$ for $\rh=1$, 
different values of $l$ and a conveniently chosen frequency $\omega=8$. (b) Here 
we plot the bound on the greybody factor for a scalar field in $3-$Brane with 
respect to varying spacetime dimensionality $n$ for $\rh=1$, different values of 
$l$ and a conveniently chosen frequency $\omega=8$. Both of the plots correspond 
to the second case $h=\sqrt{\omega^2-\mathbb{V}}$.}
 \label{fig:c2_lnon0_Bulk}
\end{figure}

where $r_{peak}$ and $\mathbb{V}_{peak}$ shows the behavior similar to the 
scalar field in the bulk. However the increment in $\mathbb{V}_{peak}$ with 
respect to the dimension $n$ is lesser compared to the bulk.
From Eqn. (\ref{eq:GBF-case2-3}) it is observed that as the value of 
$\mathbb{V}_{peak}$ increases the bound on the greybody factor decreases. Then 
the above tables imply  that the greybody factor decreases with increasing 
spacetime dimensionality.
In FIG. \ref{fig:c2_ndep_Br} one can observe that for a brane-localized scalar 
field the bound on the greybody factor decreases as the spacetime dimensionality 
$n$ increases. In the second part of the same figure we have also plotted the 
ratio of the brane to bulk bound on the greybody factor. We observed that this 
quantity is always greater than one and increases as $n$ increases to large 
values, suggesting that for large extra dimensions the black hole emit Hawking 
radiation mainly in the brane. When $l>0$ the analytical expression of 
$r_{peak}$ for a scalar field in the bulk can be given by
\begin{equation}\label{eq:rpeak_c2_ln0_Bulk}
 r_{peak} = \rh~  
\left[\frac{-\gamma_{1}+\sqrt{\gamma_{1}^2+4\gamma_{2}}}{2~(n-2)^3}\right]^{ 
-\frac{1}{ n-3}},
\end{equation}
where $\gamma_{1} = (n-1)\left[2+2l^2+2l(n-3)-n\right]$ and $\gamma_{2} = 
(n-2)^3\left[(n-2)^2+2\gamma_{1}/(n-1)\right]$, which also asserts that for a 
fixed $l$ the value of $r_{peak}$ gets nearer to the event horizon as higher 
dimensions are considered. On the other hand for a brane localized scalar field, 
when $l>0$, we have

\begin{figure}
 \includegraphics[width=0.8\linewidth]{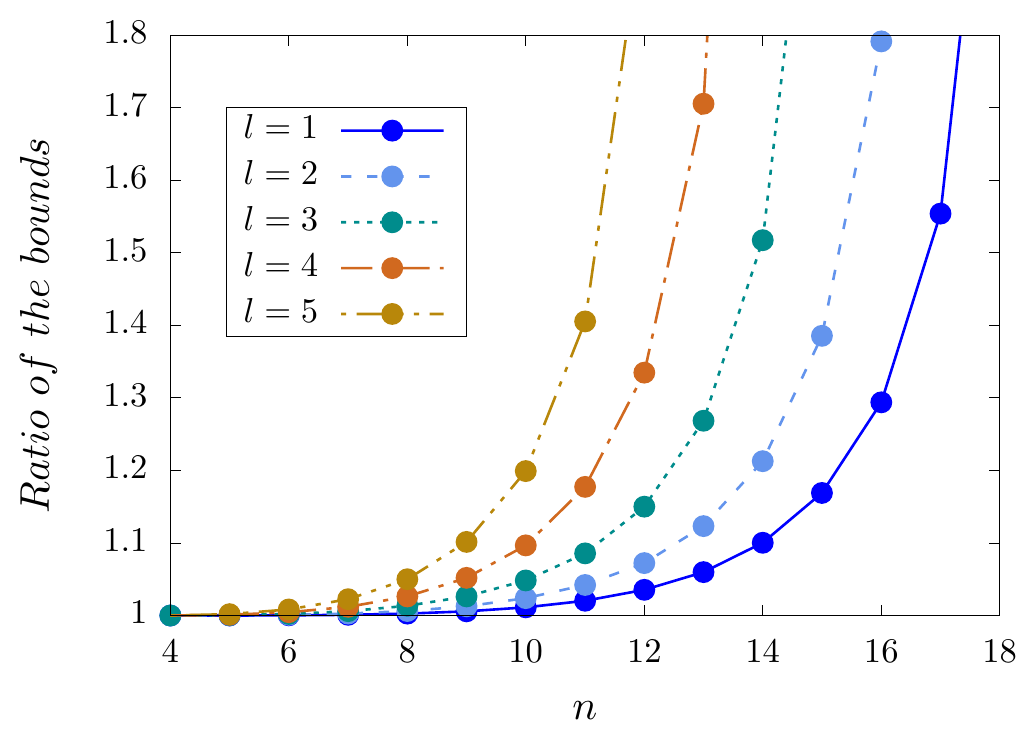}
 \caption{Here we plot the ratio of the bounds for brane to bulk localized 
scalar fields with respect to varying spacetime dimensionality $n$ for $\rh=1$, 
different values of $l$ and a conveniently chosen frequency $\omega=8$. This 
plot corresponds to the second case $h=\sqrt{\omega^2-\mathbb{V}}$.}
 \label{fig:c2_lnon0_ratio}
\end{figure}

%
\begin{equation}\label{eq:rpeak_c2_ln0_Br}
 r_{peak} = \rh~
\left[\frac{\gamma_3+\sqrt{\gamma_3^2+16l(l+1)\gamma_4}}{4~\gamma_4}\right]^{
-\frac{1}{n-3}},
\end{equation}
where $\gamma_3 = -(3+l+l^2-n)(n-1)$ and $\gamma_4 = (n-3)(n-2)$. We have 
provided a plot in FIG. \ref{fig:c2_lnon0_Bulk} depicting the bound on the 
greybody factor with respect to the spacetime dimension $n$ for $l>0$ 
corresponding to bulk and brane-localized scalar fields. It is observed that as 
$n$ increases the bound decreases even for different $l(\neq0)$. On the other 
hand from FIG. \ref{fig:c2_lnon0_ratio} it is observed that the brane to bulk 
ratio of the bound increases as the spacetime dimensionality $n$ increases. This 
further reaffirms that the quanta of the Hawking effect, with large number of 
extra dimensions, is mostly radiated in the brane.

\section{Discussion}\label{discussion}

Higher dimensional black holes provide some very interesting black hole 
spacetimes. In this work we have considered a $n-$dimensional 
Schwarzschild-Tangherlini black hole spacetime and shown that the action 
corresponding to massless minimally coupled free scalar fields both in the bulk 
and in $3-$brane can be simplified to express a $(1+1)$ dimensional flat 
spacetime in the near horizon and asymptotic regions. This particular 
characteristic allows one to express the field modes in terms of plane wave 
solutions in these regions, a feature necessary for the formulation of the 
Hawking effect. Then with the help of the near-null coordinates 
\cite{Barman:2017fzh, Barman:2017vqx, Barman:2018ina} we provided a Hamiltonian 
based derivation of the Hawking effect in the Schwarzschild-Tangherlini black 
hole spacetime. We observed that the temperature corresponding to the Hawking 
effect is dependent on the spacetime dimensionality and same for both the bulk 
and brane-localized scalar field. The second feature is obvious as the 
temperature corresponding to the Hawking effect only depends on the black hole 
horizon and for both of the cases the horizon structure is the same. However, 
the geometries outside the horizon are different, which results in different 
form of the greybody factors in bulk and in the brane. We observed that the 
bounds on the greybody factors for both bulk and brane-localized scalar field 
decrease as the spacetime dimensionality $n$ increases, a phenomenon also 
mentioned in \cite{Creek:2006ia} for bulk scalar fields. We noticed that this 
decrease is slower for the case of a brane-localized scalar field. We have also 
observed for both the bulk and brane-localized scalar field that as the 
frequency increases the bound on the greybody factor increases signifying that 
the potential barrier becomes more transparent corresponding to the field modes 
of higher frequency. These bounds from Eqn. (\ref{eq:GBF-case1-1_bulk}) and Eqn. 
(\ref{eq:GBF-case1-1_brane}), also predict that as the angular momentum quantum 
number $l$ increases the transmission of the field modes through the potential 
barrier gets depleted. We have further observed that as the spacetime 
dimensionality $n$ increases the ratio of the bounds on the greybody factor for 
brane to bulk increases and this quantity is always greater than one. This 
signifies that for a higher dimensional black hole with large extra spatial 
dimensions the Hawking emission gets specifically restricted to the brane, which 
is widely predicted in the literature \cite{Emparan:2000rs, Harris:2003eg, 
Jung:2005pk}. However, in this work we have inferred this entirely from the 
consideration of the bounds on the greybody factor. Most of the calculations 
related to these bounds can be analytically performed and they are valid in the 
entire frequency regime of the Hawking spectra for all values of the angular 
momentum quantum number $l$. We want to mention that one can also calculate 
these bounds on the greybody factor explicitly for the higher dimensional 
rotating black holes \cite{Cvetic:1997xv, Boonserm:2014fja, Jung:2005pk} and 
study the corresponding dependence on the spacetime dimensionality.

\begin{acknowledgments}
I thank Golam Mortuza Hossain, Sumanta Chakraborty and Gopal Sardar for 
discussions. I also thank Indian Institute of Science Education and Research 
Kolkata (IISER Kolkata) for supporting this work through a doctoral fellowship.
\end{acknowledgments}



\end{document}